\documentclass[aps,pra,reprint,showpacs,amssymb,amsmath,citeautoscript]{revtex4-1}
\usepackage{graphicx}

\def\be{\begin{equation}}
\def\ee{\end{equation}}
\def\ber{\begin{eqnarray}}
\def\eer{\end{eqnarray}}
\def\bern{\begin{eqnarray*}}
\def\eern{\end{eqnarray*}}

\def\rv{\mathbf{r}}

\def\jv{\mathbf{j}}
\def\Gv{\mathbf{G}}

\def\qv{\mathbf{q}}

\def\Ev{\mathbf{E}}

\def\0v{\mathbf{0}}
\def\1v{\mathbf{1}}
\def\2v{\mathbf{2}}
\def\3v{\mathbf{3}}

\begin{document}

\title{Negative static permittivity and violation of Kramers-Kronig relations in  quasi-two-dimensional crystals}
\author{V.~U.~Nazarov}
\affiliation{Research Center for Applied Sciences, Academia Sinica, Taipei 11529, Taiwan}

\begin{abstract}
We investigate the wave-vector and frequency-dependent screening of the electric field  in atomically thin
(quasi-two-dimensional) crystals.
For graphene and hexagonal boron nitride we find that, above a critical wave-vector $q_c$, the static permittivity $\varepsilon(q \! > \!q_c,\omega \! = \!0)$ becomes negative 
and the Kramers-Kronig relations  do not hold for $\varepsilon(q \! > \! q_c,\omega)$.
Thus, in quasi-two-dimensional crystals, we reveal the physical confirmation of a proposition put forward decades ago (Kirzhnits, 1976), allowing for the breakdown of Kramers-Kronig relations and for the negative static  permittivity. 
In the vicinity of the critical wave-vector, we find a giant growth of the permittivity.
Our results, obtained in the {\it ab initio} calculations using both the random-phase approximation
and the adiabatic time-dependent local-density approximation, and further confirmed with a simple slab model,
allow us to argue that the above properties, being exceptional in the three-dimensional case, are common to quasi-two-dimensional systems.
\end{abstract}

\pacs{77.22.Ch, 73.22.Pr}

\maketitle

The concept of causality plays one of the central roles in contemporary science \cite{Bohm-05}.
It is well known  that causality in the time-domain (the impossibility for an effect
to precede the cause in time) leads to the analyticity 
of a causal response-function  in a complex half-plane
in the frequency-domain, which, in
turn, leads to  Kramers-Kronig (KK) relations between the real and the imaginary parts of the  response function   \cite{Landau-60}.
  
It must be, however, recognized that the causality assumes that the response-function is applied to a cause
and it produces an effect. In the case of the longitudinal electric field
in a translationally invarient  or a periodic system,
the definition of the permittivity $\varepsilon(\qv,\omega)$ reads
$
\phi_{\text{tot}}(\qv,\omega)= \phi_{\text{ext}}(\qv,\omega)/\varepsilon(\qv,\omega),
$
where $\phi_{\text{ext}}$ and $\phi_{\text{tot}}$ are the
scalar potentials of  the externally applied  and the total  electric fields, respectively.
Since the cause is $\phi_{\text{ext}}$ and the effect is $\phi_{\text{tot}}$,  {\it not vice versa}, this is 
 $1/\varepsilon$ that is guaranteed to be causal, but not $\varepsilon$ itself \cite{Pines-66}.
Accordingly, KK relations  must be satisfied by $1/\varepsilon$, but
may or may not be satisfied by $\varepsilon$. For $|\qv|>0$, this leaves 
$\varepsilon(\qv,\omega=0)$  a freedom to be negative without violating the causality
or destroying the stability of the system \cite{Kirzhnits-76,Dolgov-81}.
If this happens, then the inverse permittivity  has zeros in the upper half of the complex $\omega$-plane,
making the permittivity itself a non-analytic function. 

In the three-dimensional world the realizations of such negative static permittivity are scarce
and  they mostly concern exotic non-crystalline systems \cite{Hansen-78,Chandra-89,Fonseca-90,Aniya-90,Yan-13-2}.
In this work we show that, above a critical wave-vector $q>q_c$ in the first Brillouin zone, the  permittivity 
$\varepsilon(\qv,\omega)$  of the quasi-two-dimensional (Q2D) systems of the  monolayer graphene 
and boron nitride is negative in the static limit.
Accordingly, KK relations for the permittivity do not hold in this case. The
inverse permittivity, on the contrary, remains causal and does satisfy KK relations.

We start by writing  the permittivity of a Q2D crystal  \cite{Nazarov-15}
(atomic units $e^2=\hbar=m_e=1$ are used throughout unless otherwise indicated)
\begin{equation}
\frac{1}{\varepsilon(\qv,\omega)} = 1+  \frac{2\pi}{q} \int\limits_{-\infty}^{\infty}  \chi_{\0v \0v}(z,z',\qv,\omega) d z d z',
\label{epsdef}
\end{equation}
where $\chi_{\Gv \Gv'}(z,z',\qv,\omega)$ is the density-response function of the system 
in the mixed, reciprocal in the system plane ($x y$) and real in the $z$-direction representation
($\Gv$ are the 2D reciprocal lattice vectors). 
\begin{figure}[h!]
\includegraphics[width= 0.8 \columnwidth, trim= 0 20 0 20, clip=true]{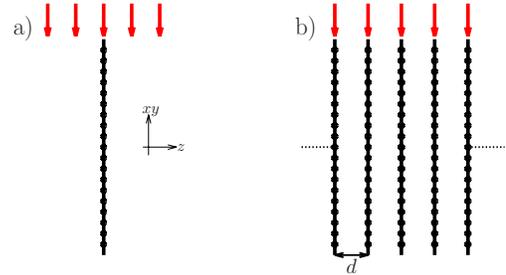} 
\caption{\label{arr} (Color online) Schematics of  2D material under an external field.
a) Q2D single-layer geometry and b)  3D super-cell geometry.}
\end{figure}

Our time-dependent density-functional theory (TDDFT) calculation of the permittivity  consists of two steps.
Since  Q2D systems lack periodicity in the $z$-direction,
it is customary to use the super-cell method \cite{Rozzi-06,Despoja-13,Nazarov-14}.
First, in the super-cell geometry, we calculate the density-response function
$\tilde{\chi}_{\Gv g, \Gv' g'}(\qv,\omega;d)$
of an auxiliary  3D system
comprised of an infinite periodic array of monolayers with the separation $d$ between  them,
as is schematized in Fig.~\ref{arr} b)
(the dependence of $\tilde{\chi}$ on $d$ is shown explicitly). 
Here $g$ are reciprocal vectors in the $z$-direction.

We have conducted  the calculation for the monolayer pristine graphene  
using  the full-potential linear augmented plane-wave (FP-LAPW) code Elk \cite{Elk}. 
The $z$-axis period $d$ of the super-cell was taken 20 a.u.
The $k$-point grid of $512\times 512 \times 1$, 30 empty bands, and  the damping parameter of 
0.002 a.u. were used in both the ground-state and the linear-response calculations. 
The former was carried out within the local-density approximation (LDA) \cite{Perdew-92} for the exchange-correlation (xc) potential, while the latter was  the random-phase approximation (RPA) one (i.e., the xc kernel $f_{xc}$ \cite{Gross-85} was set to zero).

\begin{figure}[h!]
\includegraphics[width= \columnwidth, trim= 26 0 0 0, clip=true]{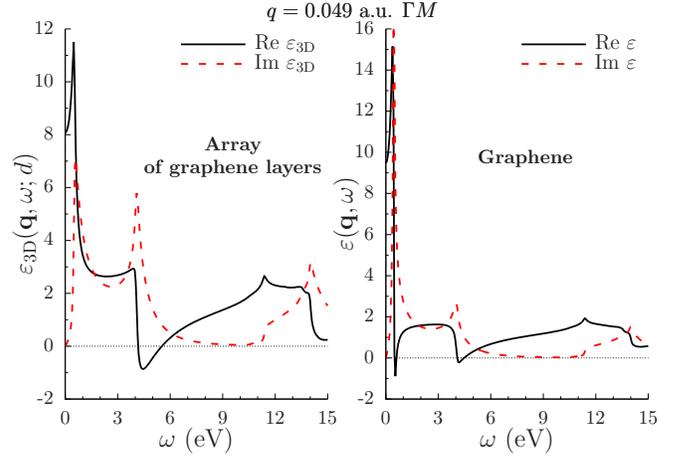}
\caption{\label{smaller} (Color online)
Left: 3D permittivity of the array of graphene layers. Right: Permittivity of a single
graphene layer. The wave-vector $q=0.049$ a.u. is below the critical value $q_c\approx 0.118$ a.u.
}
\end{figure}

\begin{figure}[h!]
\includegraphics[width= \columnwidth, trim= 26 0 0 0, clip=true]{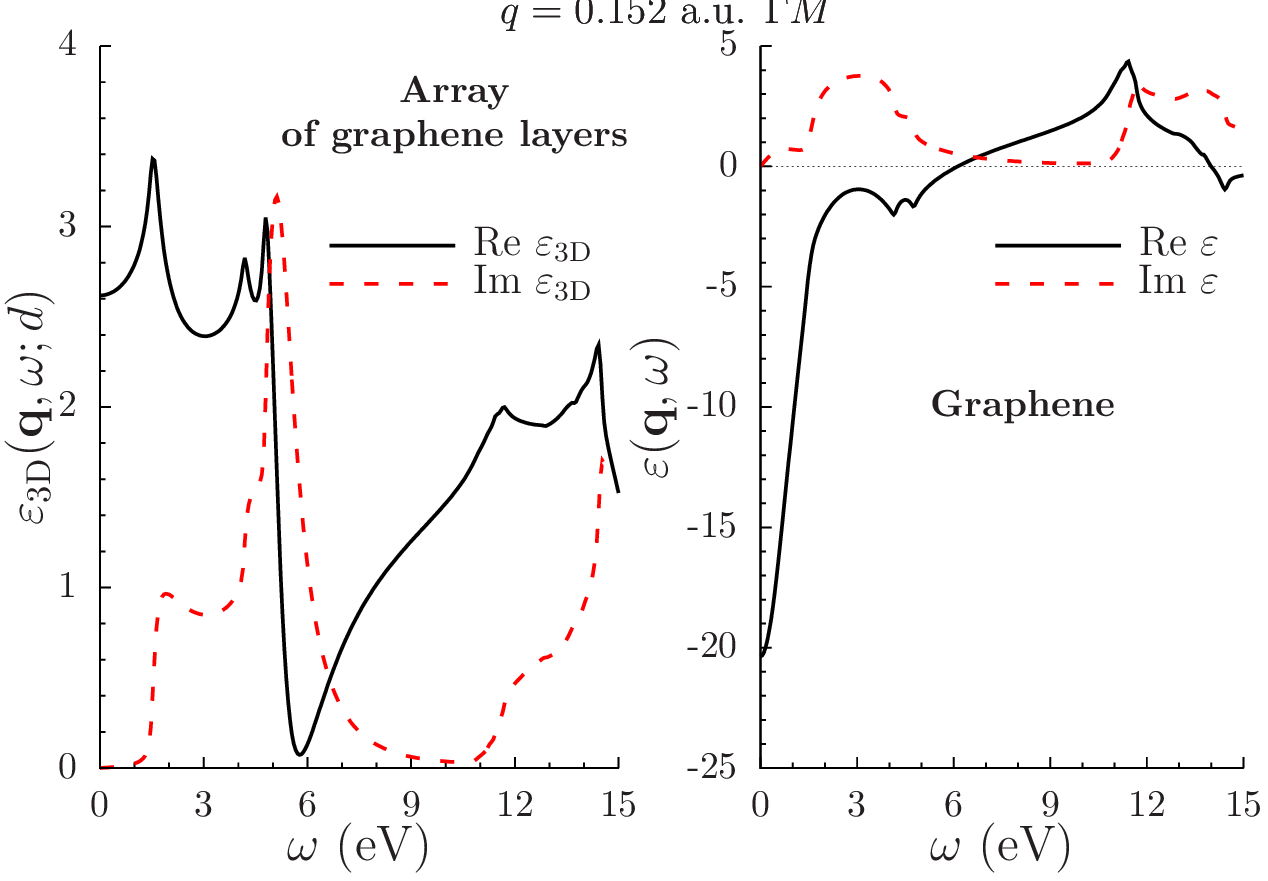}
\caption{\label{greater}  (Color online)
Left: 3D permittivity of the array of graphene layers. Right: Permittivity of a single
graphene layer. The wave-vector $q=0.152$ a.u. is above the critical value $q_c\approx 0.118$ a.u.
}
\end{figure}

Results for $\varepsilon_{\text{3D}}$, obtained through
\begin{equation}
\frac{1}{\varepsilon_{\text{3D}}(\qv,\omega; d)}=1+\frac{4\pi}{q^2} \tilde{\chi}_{\0v 0,\0v 0}(\qv,\omega;d),
\label{e3D}
\end{equation}
are presented in the left panels of Figs.~\ref{smaller} 
and \ref{greater}  for $q=0.049$ and $0.152$ a.u., respectively, along the $\Gamma M$ direction.

It is, however, known  that $\varepsilon_{\text{3D}}(\qv,\omega;d)$, calculated in the super-cell
geometry, is a quantity completely different from the permittivity $\varepsilon(\qv,\omega)$ of a single layer \cite{Rozzi-06,Despoja-13,Nazarov-14,Nazarov-15}, as can be also immediately appreciated from the $d$-dependence of the former. 
Our second step consists, therefore, in finding the density-response function $\chi$ of the single-layer system
from that of the array of those layers $\tilde{\chi}$. This can be conveniently done by virtue of the matrix relation \cite{Nazarov-15}
\begin{equation}
\chi(\qv,\omega)= \tilde{\chi}(\qv,\omega) \left[1+ C(\qv) \tilde{\chi}(\qv,\omega) \right]^{-1},
\label{fn_text}
\end{equation}
where the elements of the matrix $C$ are given by 
\begin{equation}
\begin{split}
&C_{\Gv g,  \Gv' g' }(\qv) =F_{g g'}(|\Gv+\qv|) \delta_{\Gv \Gv'}, \\
&F_{g g'}(p) \! = \! \frac{4\pi (p^2-g g')}{p d (p^2 \! + \! g^2)(p^2 \! + \! {g'}^2)} \! \cos \! \! \left[ \frac{(g \! + \! g') d}{2} \right]
\! \! (1 \! - \! e^{-p d}).
\end{split}
\label{fn_text_F}
\end{equation}
In particular, $\chi$ calculated by Eqs.~(\ref{fn_text})-(\ref{fn_text_F}) is free of the spurious inter-layer interaction,
which is present in $\tilde \chi$.

By the use of Eqs.~(\ref{fn_text})-(\ref{fn_text_F}), we find $\chi$
in the 3D reciprocal-space representation. Then, by the inverse Fourier transform to the mixed representation
and using Eq.~(\ref{epsdef}), we obtain the permittivity  $\varepsilon(\qv,\omega)$.
The latter is plotted in the right panels of Figs.~\ref{smaller} and \ref{greater}.

\begin{figure}[h!]
\includegraphics[width= \columnwidth, trim= 15 0 0 0, clip=true]{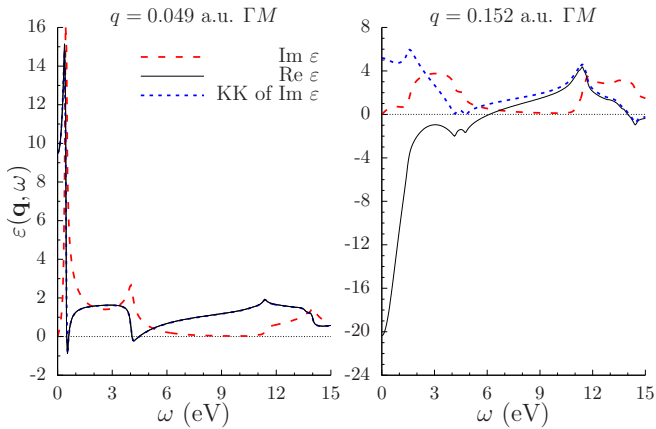}
\caption{\label{KK} (Color online)
Real and imaginary parts of the permittivity of a single-layer graphene.
Separately is plotted the real part obtained from the imaginary part by the use of  KK relation. 
Left: $q< q_c$, KK relation holds. Right: $q> q_c$, KK relation does not hold.
}
\end{figure}

A striking feature in Fig.~\ref{greater}, right panel, is that $\varepsilon(\qv,\omega  =  0)$ is negative
and, since $\text{Im} \, \varepsilon(\qv,\omega \ge 0) \ge 0$, the permittivity of graphene
does not satisfy Kramers-Kronig relations \cite{Landau-60}
\begin{equation}
\begin{split}
&\text{Re} \, R(\omega) = R(\infty) + \frac{2}{\pi} \mathcal{P} 
\int\limits_0^\infty \frac{\omega' \text{Im} \, R(\omega')}{\omega'^2-\omega^2} d\omega', \\
&\text{Im} \, R(\omega) = - \frac{2 \omega}{\pi} \mathcal{P} 
\int\limits_0^\infty \frac{\text{Re} \, R(\omega') - R(\infty)}{\omega'^2-\omega^2} d\omega',
\end{split}
\label{KKr}
\end{equation}
$\mathcal{P}$ denoting the principal value of the integrals, with $R(\omega)=\varepsilon(\qv,\omega)$.
This fact is further illustrated in Fig.~\ref{KK},
where the real part of $\varepsilon(\qv,\omega)$ is compared with the KK transform of 
its imaginary part: The two functions coincide at $q<q_c$ (left panel), but they are largely different at $q>q_c$ (right panel).
We found the critical wave-vector for graphene to be $q_c \approx 0.118$ a.u. ($0.223$ \AA$^{-1}$). 
\footnote{Generally speaking, $q_c$ is different in different directions, however, it is practically isotropic in graphene.} 
On the other hand, it can be seen in Figs.~\ref{smaller} and \ref{greater}, left panels, that the array system 
has a positive static permittivity, which cannot be otherwise for a 3D periodic system within RPA \cite{Dolgov-81}.

Since KK relations are not satisfied  by $\varepsilon(\qv,\omega)$, 
the latter must have a singularity in the complex 
$\omega$ upper half-plane. The singularity can only be a pole at  $\omega=\omega_s$ satisfying 
\begin{equation}
\frac{1}{\varepsilon(\qv,\omega_s)}=0.
\label{equ}
\end{equation}
Considering that  (a) $1/\varepsilon(\qv,\omega)$ is a real continuous function on the positive imaginary axis of the $\omega$-plane;
(b) $1/\varepsilon(\qv,\omega=0)<0$ for $q>q_c$; and (c) $1/\varepsilon(\qv,\omega=i \infty)=1$,
we conclude that, at $q>q_c$, there exists a point $\omega_s$ on the positive imaginary axis of the $\omega$-plane
which satisfies  Eq.~(\ref{equ}).
To find this point, we write by virtue of Cauchy's integral formula
\begin{equation}
\frac{1}{\varepsilon(\qv,i u)} -1 =  \frac{1}{2\pi i} \int\limits_{-\infty}^{\infty}
\frac{\frac{1}{\varepsilon(\qv,\omega')}-1}{\omega'-i u} d\omega'.
\label{Cau}
\end{equation}
Expanding  the complex inverse pemittivity in the right-hand side of Eq.~(\ref{Cau}) via its real and imaginary parts and using the parity properties of those functions, we can write
\begin{equation}
\begin{split}
\frac{1}{\varepsilon(\qv,i u)}  =   1 
 &+  \frac{1}{\pi}  \int\limits_{0}^{\infty} u
 \frac{  \text{Re}\,\frac{1}{\varepsilon(\qv,\omega')}-1 }{{\omega'}^2  +  u^2 }  d\omega' \\ 
 &+  \frac{1}{\pi}  \int\limits_{0}^{\infty} \omega'
 \frac{\text{Im}\, \frac{1}{\varepsilon(\qv,\omega')}}{{\omega'}^2  +  u^2 }  d\omega'.
\end{split}
\end{equation}
Further, the equality of the second term on the right-hand side to the third one can be easily proven with the use
of the KK relations for $1/\varepsilon(\qv,\omega)$. We then have 
\begin{equation}
\frac{1}{\varepsilon(\qv,i u)}  =   1  +  \frac{2}{\pi}  \int\limits_{0}^{\infty}
\omega' \frac{\text{Im}\, \frac{1}{\varepsilon(\qv,\omega')}}{{\omega'}^2  +  u^2 }  d\omega'.
 \label{iKK}
\end{equation}

\begin{figure}[h!]
\includegraphics[width= \columnwidth, trim= 20 0 0 0, clip=true]{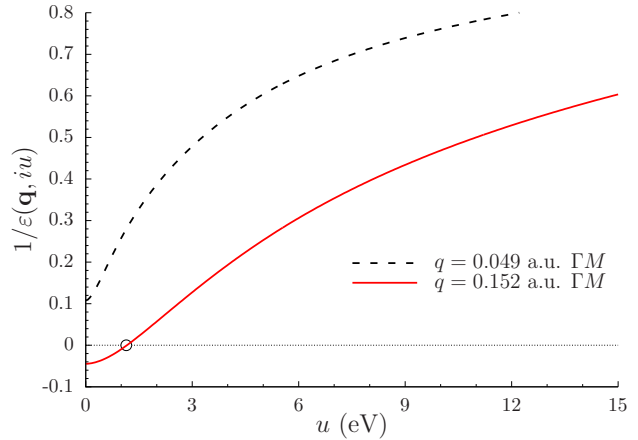}
\caption{\label{foi} (Color online)
Inverse permittivity of a single graphene layer as a function of the imaginary frequency.
At $q<q_c$ (black dashed curve), there is no zero (the permittivity is analytic in the upper complex $\omega$-plane).
At $q>q_c$ (red solid curve), the inverse permittivity has a zero (indicated by a circle). 
Accordingly, the permittivity has a pole at this frequency.
}
\end{figure}

We use Eq.~(\ref{iKK}) to calculate the inverse permittivity on  the positive imaginary $\omega$-axis from our results for it on the real axis. 
In Fig.~\ref{foi}, this is plotted for the two wave-vectors, below and above the critical value.
Above the critical wave-vector, the inverse permittivity crosses zero (indicated by a circle in Fig.~\ref{foi}),
which does not happen below the critical wave-vector. 
In  Appendix~\ref{low}, 
we compare our results with the analytical ones known in the low-$q$ regime \cite{Dobson-06,Wunsch-06}.

We can  gain a further insight into the situation using the approximate analytical relation between the Q2D permittivity
of a single layer and the 3D permittivity of the array of those layers 
\begin{equation}
\frac{1}{\varepsilon(\qv,\omega)}=
1+\frac{1}{2} \frac{1}{ \frac{1}{\left[ \frac{1}{\varepsilon_{\text{3D}}(\qv,\omega;d)} -1 \right] q d} +
\frac{1}{e^{q d}-1}} ,
\label{eps2d3d}
\end{equation}
rather than with the 'exact' numerical solution of Eqs.~(\ref{fn_text})-(\ref{fn_text_F}).
Equation (\ref{eps2d3d}), derived in Ref.~\onlinecite{Nazarov-14}, is a good approximation  at $q$ far from the critical value from the both sides,
as we demonstrate below in Fig.~\ref{exap}. 
Solving Eq.~(\ref{eps2d3d}) with respect to $\varepsilon_{\text{3D}}$, we find that $1/\varepsilon$ is zero if
\begin{equation}
\varepsilon_{\text{3D}}(\qv,\omega;d)= 1-\frac{2}{q d} \left(   \frac{e^{q d}-1}{e^{q d}+1}\right).
\label{zzz}
\end{equation}
\begin{figure}[h!]
\includegraphics[width= \columnwidth, trim=30 0 0 0, clip=true]{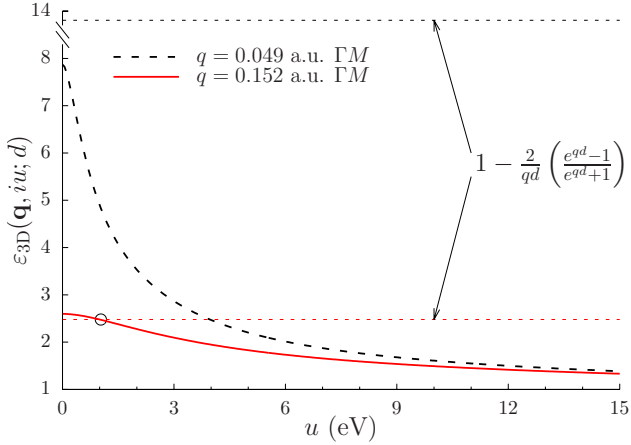}
\caption{\label{foie} (Color online)
The permittivity of an array of graphene layers as a function of the imaginary frequency.
Horizontal dashed lines show the values of $\varepsilon_{\text{3D}}$ at which, by Eq.~(\ref{zzz}),
the permittivity of a single-layer graphene may become singular. This never happens
at $q<q_c$ (black dashed curve), while this happens
at $q>q_c$ (red solid curve).
}
\end{figure}

In Fig.~\ref{foie}, we plot $\varepsilon_{\text{3D}}(\qv,\omega;d)$ along the positive
imaginary $\omega$. Although $\varepsilon_{\text{3D}}(\qv,\omega;d)$ is analytic in the upper complex $\omega$-plane at 
all values of $\qv$, it gives rise to a zero in $1/\varepsilon$ (a pole in $\varepsilon$)
when the condition (\ref{zzz}) is met. In Fig.~\ref{foie} this is shown as an intersection, 
in the case of $q>q_c$, with the straight horizontal line
representing the right-hand side of Eq.~(\ref{zzz}).

\begin{figure}[h!]
\includegraphics[width= \columnwidth, trim= 15 0 0 0, clip=true]{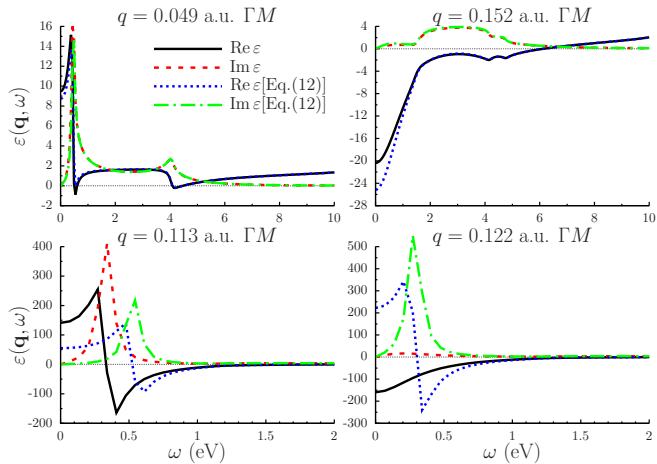}
\caption{\label{exap} (Color online)
The permittivity of single-layer graphene calculated with the 'exact' numerical procedure using Eqs.~(\ref{fn_text}), (\ref{fn_text_F}), and (\ref{epsdef}), and the approximate analytical Eq.~(\ref{eps2d3d}). 
Upper panels: The wave-vector is {\it well} below (left) and  above (right) the critical value $q_c \approx 0.118$ a.u. Lower panels: The wave-vector is {\it slightly} below (left) and  above (right) $q_c$.
}
\end{figure}

Figure~\ref{exap} is presented in support of the fact that the  permittivity obtained through
the 'exact' numerical procedure via Eqs.~(\ref{fn_text}), (\ref{fn_text_F}), and
(\ref{epsdef}) can be accurately approximated by the simple analytical formula of Eq.~(\ref{eps2d3d}), if $q$ is
sufficiently below or above the critical wave-vector (upper panels). On the contrary,  
the same comparison done for the wave-vector slightly below and above the critical value (lower panels), reveals  the complete
inapplicability of the approximate formula (\ref{eps2d3d}) in the vicinity of the critical wave-vector.
Moreover, a giant increase in the absolute value of the permittivity occurs close to the critical wave-vector.

\begin{figure}[h!]
\includegraphics[width= \columnwidth, trim= 15 0 0 0, clip=true]{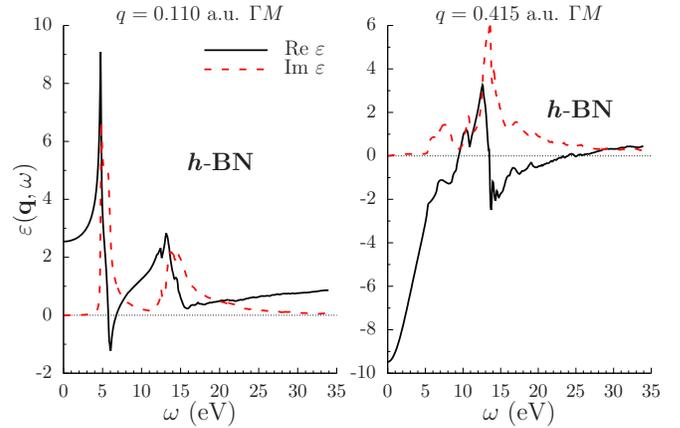}
\caption{\label{BN} (Color online)
The permittivity of 2D hexagonal boron nitride below (left) and above (right) the critical wave-vector $q_c\approx 0.323$ a.u.
}
\end{figure}

Graphene is known to be a semi-metal, possessing  a remarkable property of Dirac's cones touching in the $K$-point 
of its band-structure \cite{Neto-09}. 
A  natural question arises whether the negative static permittivity and the violation of KK relations in graphene are in any way related to the Dirac's cones in this material. To answer this,
in Fig.~\ref{BN} we present results for the permittivity of  the hexagonal  boron nitride ($h$-$\mathrm{BN}$), 
known to be an insulator \cite{Lin-12}.
Similar to graphene, above a critical wave-vector $q_c\approx 0.323$ a.u. ($0.610$ \AA$^{-1}$) (right panel of Fig.~\ref{BN}),
the permittivity of $\mathrm{BN}$ does not satisfy KK relations, while having a negative static limit. 
Furthermore, in Appendix~\ref{slabs}, we demonstrate that a simple local model
of a metallic slab in vaccum  supports the negative static permittivity at larger wave-vectors.
This  shows that
the negative static permittivity and the breakdown of KK relations is a rather general property 
common to Q2D systems.

Importantly, in perfect 3D crystals the negative static permittivity is only possible
due to the dynamic xc effects in the electronic response \cite{Dolgov-81}.
Since our results for Q2D crystals are obtained within the RPA, i.e., neglecting the
xc effects, and the negative static permittivity occurs possible, the situation is fundamentally different 
with Q2D crystals: This is the finite but microscopic thickness of the crystal, 
which is also the break of the periodicity in one dimension, that makes the negative static permittivity possible.
Nonetheless, static and dynamic many-body effects play an important part in Q2D crystals \cite{Yang-09,Kotov-12},
which have not been accounted for in the present study. 
In  Appendix~\ref{atdlda}, we show that the the inclusion of the xc kernel $f_{xc}$ 
on the level of 
the adiabatic time-dependent local-density
approximation (ATDLDA) \cite{Zangwill-80,Runge-84,Gross-85} does not lead to a significant change in the results.
The inclusion of the same effects within TDDFT with more elaborate $f_{xc}$, e.g., following the schemes known in the 3D case
\cite{Botti-04,Nazarov-11}, presents a challenge in the case of Q2D systems.
Furthermore, for Q2D crystals supported on substrates, the interaction with the latter strongly influences
the excitation processes \cite{Politano-14}, which is also a demanding problem to be addressed in the future.

For the accurate interpretation  of the results, 
it is necessary to keep in mind  the exact meaning of the permittivity (\ref{epsdef}) of a Q2D crystal. 
This definition  is given  in two steps:\cite{Nazarov-15}
First, the 2D conductivity $\sigma^{\text{ext}}_{\text{2D}}$ {\em with respect to the external field}  is introduced
\begin{equation}
\jv_{\text{2D}}(\qv,\omega)= \sigma^{\text{ext}}_{\text{2D}}(\qv,\omega) \Ev^{\text{ext}}(\qv,\omega),
\end{equation}
where $\Ev^{\text{ext}}(\qv,\omega)$ is the uniform in the $z$-direction external electric field,
and $\jv_{\text{2D}}(\qv,\omega)$ is the 3D current-density {\em integrated in the $z$-direction}
and averaged over the unit cell in the $xy$-plane.
Secondly, the permittivity of a Q2D crystal is defined by the  relation
\begin{equation}
\frac{1}{\varepsilon(\qv,\omega)}=1+\frac{2\pi q}{i \omega} \sigma^{\text{ext}}_{\text{2D}}(\qv,\omega),
\label{epsdef2}
\end{equation}
rigorously valid for a strictly 2D system. 
The final justification of Eq.~(\ref{epsdef2}) is that with this definition the usual formula for the energy dissipation
\begin{equation}
Q(\qv,\omega)= - \frac{\omega }{4\pi q} |\Ev_{\text{ext}}(\qv,\omega)|^2 \, {\rm Im}\, \frac{1}{\varepsilon(\qv,\omega)} 
\label{QQQ}
\end{equation}
holds for a Q2D crystal exactly. However, as detailed in Ref.~\onlinecite{Nazarov-15}, the Q2D permittivity cannot be attributed
the meaning of the coefficient of proportionality between the external and the total fields
and, hence, Eq.~(\ref{QQQ}) cannot be rewritten in terms of $\Ev_{\text{tot}}$ and ${\rm Im}\, \varepsilon$.
We note that these complications call for the particular caution
in the consideration of the negative static permittivity in the context of 
the 2D superconductivity \cite{Kirzhnits-76,Dolgov-81}.
The behavior of the total field along the $z$-direction in graphene is further discussed in 
Appendix~Sec.~\ref{distr}.

In conclusions, we have established the violation of Kramers-Kronig relations by the wave-vector and frequency dependent permittivity
of quasi-two-dimensional  graphene and boron nitride above a critical magnitude of the wave-vector,
and the static permittivity was found negative in this case.
The mechanism for the negative static permittivity was shown conceptually different from that in the 3D case:
It is due to the system finite microscopic thickness rather than to the exchange-correlation effects.
Our findings  suggest the fundamental differences of the screening 
and the electronic excitation processes in quasi-2D
crystals as compared with both 3D and purely 2D systems.
It is, however, discussed that further work is required to consider the present results in the context of the 2D superconductivity.

\acknowledgments
I thank Guang-Yu Guo for valuable discussions.
Support from the Ministry of Science and Technology, Taiwan, Grants 103-2112-M-001-007 and 104-2112-M-001-007, is acknowledged. 
                                       
%\bibliography{ref}

%merlin.mbs apsrev4-1.bst 2010-07-25 4.21a (PWD, AO, DPC) hacked
%Control: key (0)
%Control: author (8) initials jnrlst
%Control: editor formatted (1) identically to author
%Control: production of article title (-1) disabled
%Control: page (0) single
%Control: year (1) truncated
%Control: production of eprint (0) enabled
%

\onecolumngrid

\appendix

\section{Low-$Q$  permittivity of graphene}
\label{low}
\begin{figure}[h!]
\includegraphics[width= 0.5 \columnwidth, trim= 0 0 0 0, clip=true]{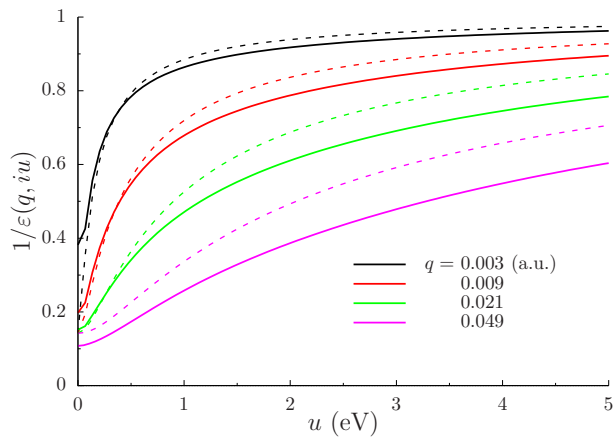}
\caption{\label{foidob} Inverse permittivity of graphene as a function of the imaginary frequency calculated
within the framework of this paper (solid lines) and its analytical long-wave behaviour (\ref{q0})
within a 2D model \cite{Dobson-06,Wunsch-06} (dashed lines).
}
\end{figure}

At small $q$, the permittivity of graphene, 
thought as a strictly 2D system, can be written as a function of the imaginary frequency 
as \cite{Dobson-06,Wunsch-06} (the density-response function of Eq.~(2) in Ref.~\onlinecite{Dobson-06} must be two times less)
\begin{equation}
\varepsilon(q,i u)=1+\frac{\pi q}{2 \sqrt{u^2+(q v_f)^2}},
\label{q0}
\end{equation}
where $v_f\approx$ 0.26 a.u. \cite{Dobson-06} is the Fermi velocity.
In Fig.~\ref{foidob} we compare our first-principles results for graphene as a Q2D system to those from
Eq.~(\ref{q0}). The conclusions are as follows:
\begin{enumerate}
\item
At very small wave-vector ($q=0.003$ a.u.) the two calculations are in good agreement except for $u$ at and very close to zero. 
The latter disagreement at very small $q$ and $u$ (or $\omega$, in the real-frequency calculation)
is due to a the principle difficulty in achieving a good accuracy
in the numerical calculation with finite $k$-grid and damping in the very vicinity of the Dirac point.
In this regime results obtained through Eq.~(\ref{q0}) can be thought superior to the {\it ab initio} ones.
See also discussion of Fig.~6 in Ref.~\onlinecite{Nazarov-15}.
\item
Small but not too small wave-vector ($q=0.009$ and $0.021$ a.u.). The agreement between the two calculations
is very good in the low $u$ range. There is, however, no reason for them to agree at higher $u$,
since the analytical formula is built on the two-bands model, while the numerical calculation
uses the realistic band-structure (30 bands for this calculation) (see also  Fig.~6 in Ref.~\onlinecite{Nazarov-15}
to illustrate the same point in the real-frequency calculation). 
\item
Larger wave-vector ($q=0.049$ a.u.). The two calculations disagree, the small-$q$ expansion being not relevant any more.
\end{enumerate} 
In both regimes 2 and 3 the numerical results are superior to the analytical ones.

%\newpage

\section{Slab with a local constituent permittivity: illustrative phenomenological model}
\label{slabs}
\begin{figure}[h!]
\includegraphics[width= 0.35 \columnwidth, trim= 20 20 20 20, clip=true]{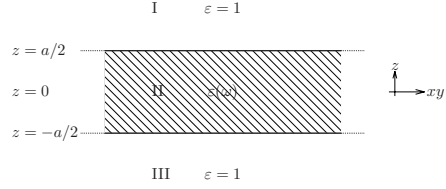}
\caption{\label{slabg} A slab with the local permittivity $\varepsilon(\omega)$ and the thickness $a$ surrounded by vacuum.
}
\end{figure}
In order to illustrate the general nature of the negative static permittivity and, consequently,
of the violation of KK relations in Q2D systems, in this section we consider a simple model
of a slab of the thickness $a$ comprised of a uniform medium with the local permittivity $\varepsilon(\omega)$,
as shown in Fig.~\ref{slabg}.
Let the externally applied potential be
\begin{equation}
\phi_{\text{ext}}(\rv,\omega)= e^{i(\qv\cdot \rv-\omega t)},
\end{equation}
where $\qv$ lies in the $xy$ plane. Then the total potential in the regions I, II, and III can be written as
\begin{equation}
\phi(z)=
\left\{       
\begin{array}{ll}
\text{I}  &  1+A_1 e^{-q z}, \\
\text{II} &  \frac{1}{\varepsilon}+A_2 e^{-q z} + B_2 e^{q z}, \\
\text{III} & 1+B_3 e^{q z},
\end{array}
\right.
\label{123}
\end{equation}
where $A_1$, $A_2$, $B_2$, and $B_3$ are constants to be found from the boundary
conditions
\begin{equation}
\begin{split}
1+A_1 e^{-q a/2}= \frac{1}{\varepsilon}+A_2 e^{-q a/2} + B_2 e^{q a/2}, \\
1+B_3 e^{-q a/2}= \frac{1}{\varepsilon}+A_2 e^{q a/2} + B_2 e^{-q a/2}, \\
- A_1 e^{-q a/2}= \varepsilon \left[- A_2 e^{-q a/2} + B_2 e^{q a/2} \right], \\
B_3 e^{-q a/2}= \varepsilon \left[ -A_2 e^{q a/2} + B_2 e^{-q a/2} \right] ,
\end{split}
\label{boundary}
\end{equation}
where the first two are due to the continuity of $\phi(z)$ at the interfaces
and the last two to the continuity of the $z$-component of the displacement vector $D_z= - \varepsilon(z) \frac{d \phi(z)}{d z}$.
The solution of Eqs.~(\ref{boundary}) is
\begin{equation}
\begin{split}
&A_2=B_2= \frac{\varepsilon-1}{\varepsilon} \frac{e^{a q/2}}{1+e^{a q}+\varepsilon (e^{a q}-1)},\\
&A_1=B_3= \varepsilon \left( 1-e^{a q} \right) A_2 . 
\end{split}
\label{AB}
\end{equation}

The induced charge-density can be found as
\begin{equation}
\rho(z)= -\frac{1}{4\pi} \left( \frac{d^2}{d z^2} -q^2 \right)  \left[ \phi(z)-\phi_{\text{ext}}(z) \right].
\end{equation}
Therefore, using Eqs.~(\ref{123}) and $\phi_{\text{ext}}(z)=1$, we have
\begin{equation}
\begin{split}
\rho(z)= \frac{q}{4\pi} \left\{ q \left( \frac{1}{\varepsilon}-1 \right) \Theta\left(\frac{a}{2}-z\right) \Theta\left(\frac{a}{2}+z\right)
+ \delta\left(z-\frac{a}{2}\right)\left[  \left(A_1-A_2\right) e^{-q a/2} +B_2 e^{q a/2} \right] \right. \\ \left.
+ \delta\left(z+\frac{a}{2}\right)\left[  \left(B_3-B_2\right) e^{-q a/2} +A_2 e^{q a/2} \right] \right\},
\end{split}
\end{equation}
where $\Theta(x)$ is the Heaviside step-function. Then
\begin{equation}
\int \chi(z,z') d z d z'= \int \rho(z) d z = \frac{q^2 a}{4\pi} \left( \frac{1}{\varepsilon}-1 \right)
+\frac{q}{2\pi} \left[  A_1 e^{-q a/2}+2 A_2 \sinh \left(q a/2 \right) \right],
\end{equation}
where Eqs.~(\ref{AB}) were used. Finally, using Eq.~(\ref{epsdef}), we have
\begin{equation}
\frac{1}{\varepsilon(q,\omega)}=
1+\frac{q a}{2} \left[ \frac{1}{\varepsilon(\omega)}-1 \right]
+   A_1(q,\omega) e^{-q a/2}+2 A_2(q,\omega) \sinh \left(q a/2 \right),
\label{se}
\end{equation}
where we have restored the explicit arguments of the functions.

\begin{figure}[h!]
\includegraphics[width= 0.5 \columnwidth, trim= 15 0 0 0, clip=true]{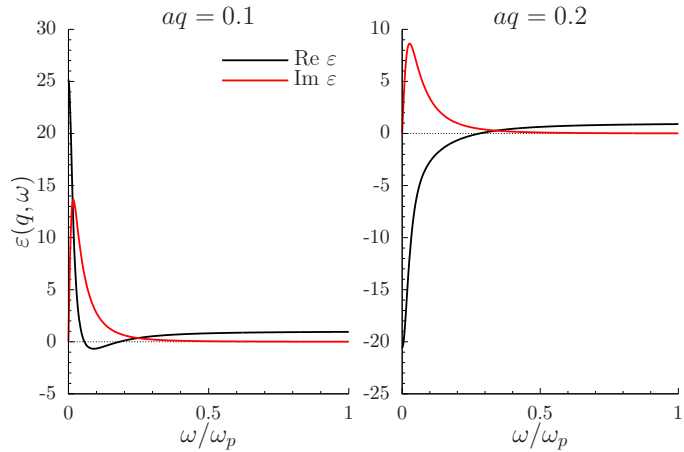}
\caption{\label{slab}
The permittivity of the metallic slab model below (left) and above (right)
the critical wave-number.
}
\end{figure}

In Fig.~\ref{slab} we plot the permittivity of Eq.~(\ref{se}) for a metallic slab
with the Drude constituent permittivity 
\begin{equation}
\varepsilon(\omega)=1-\frac{\omega_p^2}{(\omega+i \eta)^2}
\label{Drude}
\end{equation}
with $\eta/\omega_p=0.067$. Similar to the first-principles calculations for Q2D crystals,
the model system yields the negative static permittivity at lager wave-vectors (right panel of Fig.~\ref{slab}).
Finally, by the direct substitution of Eqs.~(\ref{AB}) and (\ref{Drude}) into Eq.~(\ref{se}), it is easy to show that, in the limit of the 2D electron gas ($a\rightarrow 0$, $\omega_p^2 a=4\pi n_{\text{3D}} a \rightarrow 4\pi n_{\text{2D}}$), the permittivity (\ref{se}) reduces to
\begin{equation}
\varepsilon(q,\omega) \rightarrow 1-\frac{2\pi n_{\text{2D}} q}{(\omega+ i\eta)^2},
\end{equation}
which is a standard result for the 2D electron gas in the long-wave limit,
with which, the negative static permittivity is not, of course, possible.

%\newpage

\section{Beyond RPA: Adiabatic time-dependent local-density approximation}
\label{atdlda}
\begin{figure}[h!]
\includegraphics[width= 0.5 \columnwidth, trim= 15 0 0 0, clip=true]{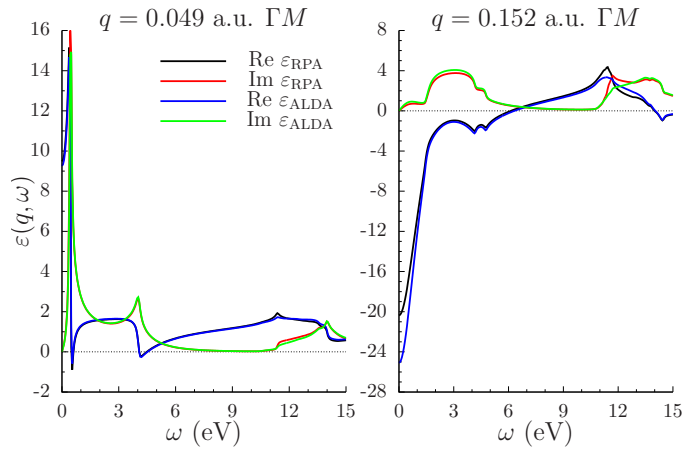}
\caption{\label{alda}
Comparison of the permittivities of a single-layer graphene within RPA (black and red lines)
and ATDLDA (blue and green lines). 
Left: $q< q_c$. Right: $q> q_c$.}
\end{figure}

In order to go beyond RPA, we have conducted calculations using the xc kernel $f_{xc}$ at the level of ATDLDA.
Results are presented in Fig.~\ref{alda}, showing no significant difference compared with RPA.
It must be noted that for TDDFT as applied to Q2D crystals, ATDLDA is the current state-of-the-art
in accounting for the dynamic xc effects. Indeed, more elaborate kernels \cite{Botti-04,Nazarov-11},
developed for 3D crystals, are not applicable to the Q2D case.

\vspace{-0.35 cm}
\section{Distribution of the charge-density and the total potential in the $Z$-direction}
\label{distr}
\begin{figure}[h!]
\includegraphics[width= 0.5 \columnwidth, trim= 40 0 10 0, clip=true]{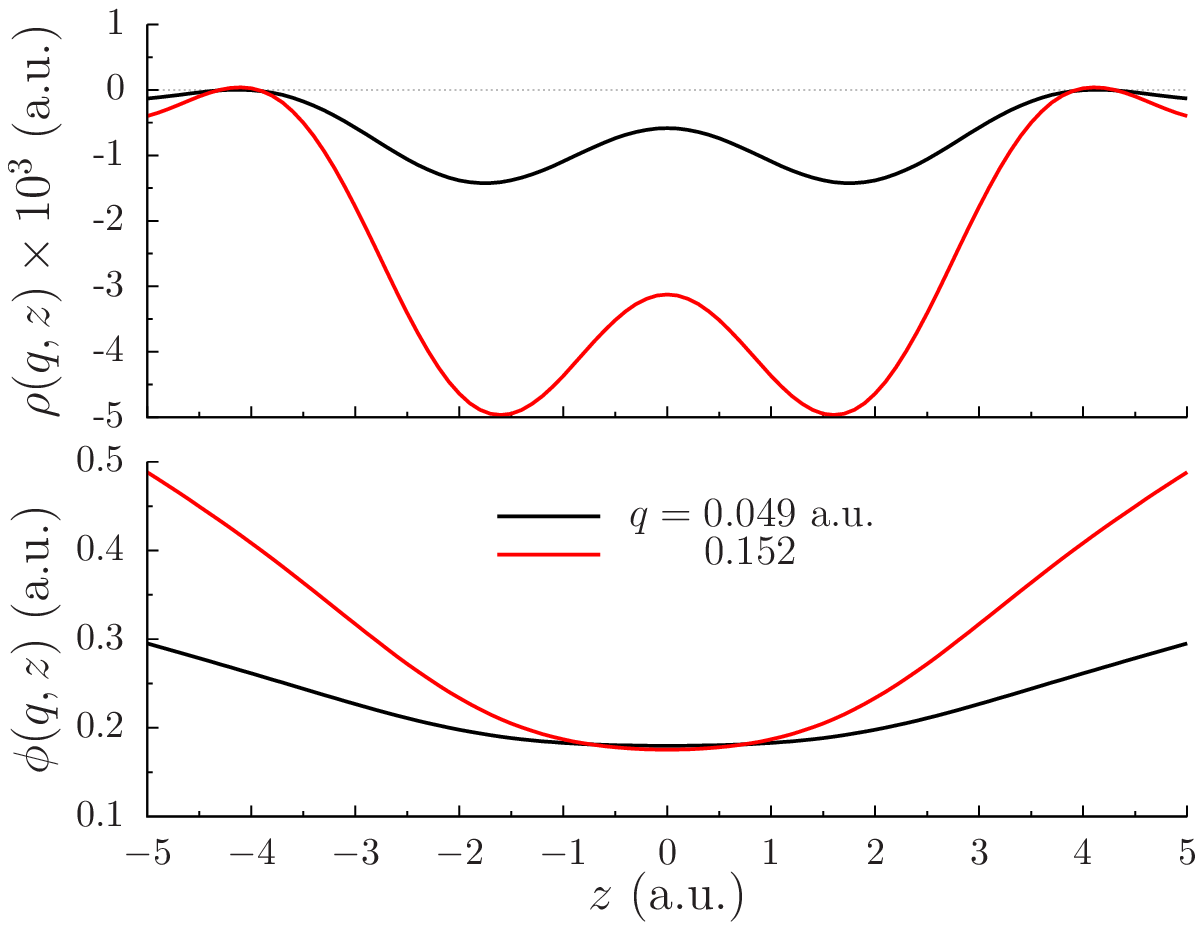}
\caption{\label{rofi}
The $z$-distribution of the charge-density (upper panel) and total potential (lower panel)
at two values of the wave-vector. The static ($\omega=0$), uniform in the $z$-direction, unity-amplitude external potential is applied. The averaging in the $xy$-plane has been performed.
}
\end{figure}

In Fig.~\ref{rofi} we plot the charge-density $\rho(q,z)$ induced in graphene and the 
corresponding total potential $\phi(q,z)$ in response to the static,  uniform in the $z$-direction, unity-amplitude external potential for the two values of the in-plane wave-vector, below and above 
the critical value $q_c$. An important feature seen from the lower panel is that,
although a strong screening occurs inside the graphene layer, the direction
of the total field does not change sign even for $q>q_c$. This prevents us from
directly relating the negative static permittivity in Q2D crystals to the 
2D superconductivity in these systems.

\end{document}